\documentclass[pra,twocolumn]{revtex4}

\usepackage[dvips]{graphicx}%
\usepackage{bm,color}
\usepackage{amsmath,amssymb}
\newcommand{\braket}[2]{\langle #1 | #2 \rangle}
\newcommand{\ketbra}[2]{\ket{#1}\bra{#2}} 
\newcommand{\ket}[1]{\left |  #1 \right \rangle}
\newcommand{\bra}[1]{ \left \langle #1  \right |}

\def \tr{{\textrm {Tr}}}

\begin{document}
\title{Simple proof of the quantum benchmark fidelity \\ for continuous-variable quantum devices} %

\author{Ryo Namiki}\affiliation{Department of Physics, Graduate School of Science, Kyoto University, Kyoto 606-8502, Japan}

\date{April 6, 2011}
\begin{abstract} 
An experimental success criterion for continuous-variable quantum teleportation and memories is to surpass a limit of the average fidelity achieved by the classical measure-and-prepare schemes with respect to a Gaussian distributed set of coherent states. We present an alternative proof of the classical limit  based on the familiar notions of the state-channel duality and the partial transposition. The present method enables us to produce a quantum-domain criterion associated with  a given set of measured fidelities. 
 \end{abstract}

\maketitle



In order to realize quantum information processing \cite{NC00}, a central challenge is 
to establish reliable quantum channels to transmit and storage quantum states faithfully. For a given experimental implementation of a quantum channel, it is natural to ask whether or not its performance originates from quantum coherence. This question is vital to assert the success of an experimental quantum teleportation \cite{Benn93} since it transmits quantum states by consuming quantum entanglement and has to maintain better fidelity of transmission beyond the classical transmission without entanglement \cite{Pop94,Mass95,Horo99}. In the present, the framework to prove the effect of entanglement can be applied for a wide class of experiments including the processes of quantum memory \cite{RMP82} and quantum key distribution \cite{rmp-qkd}. Associated with the increase of activity in experimental researches, there has been a growing interest in  producing more practical and accessible settings for the proof of entanglement  \cite{Bra00,Ham05,namiki07,Has10,namiki08,Rig06,Takano08,Has08,Fuc03,Cal09}.

A central notion to demonstrate quantum advantage 
 over the classical processes is to outperform all \textit{classical measure-and-prepare} (MP) schemes \cite{Bra00,Ham05,namiki07,Has10,namiki08,Cal09}. A classical MP scheme is an entanglement breaking (EB) channel which breaks possible entanglement shared between the system being subject to the process and any other system \cite{16}. If a process is incompatible with any EB channel, one can find an entangled state whose inseparability survives after the entangled subsystem is subject to the process. In this case we call the process is in \textit{quantum domain}.  A natural figure of merit to measure the performance of the process is  an average of the fidelities between the ideal output (\textit{target}) states and actual output states of the process over a set of input states with a certain prior probability distribution \cite{namiki07,namiki08}. The classical limit of the average fidelity achieved by the classical MP schemes is called the \textit{quantum benchmark} fidelity. Surpassing this fidelity limit is the proof of the entanglement and basic success criterion of the experiment for implementing quantum devices \cite{Furusawa98,Jul04,Lob09}. %

In quantum optics and continuous-variable quantum information processing \cite{CV-RMP}, the coherent state 
is one of the most accessible quantum states, and it is natural to test the device  by the input of coherent states. It is theoretically simple to determine the classical limit assuming the uniform distribution of coherent states, however,
 neither testing the input-output relation for every coherent state nor assuming the displacement covariant property for the real device is feasible.
 Hence, a Gaussian distribution has been employed to observe the performance on a flat distribution over a feasible amount of phase-space displacement \cite{Bra00,Ham05}. The value of quantum benchmark fidelity  with respect to the Gaussian distributed set of coherent states  had been conjectured \cite{Bra00}, and this conjecture was proven in \cite{Ham05}.   After the rigorous proof \cite{Ham05}, the classical limit fidelity for a class of non-unit-gain tasks is derived  in order to deal with highly lossy processes, such as, a long distance transmission channel and a quantum memory process with a longer storage time \cite{namiki07}. The proof \cite{Ham05} has also been utilized in the problem of the non-locality without entanglement \cite{Nis07}.   In view of these general importance, it would be insightful to find a different way to reach the fundamental benchmark.

In this report, we present an alternative proof of the quantum benchmark fidelity for continuous-variable quantum devices with respect to the transformation of Gaussian distributed set of coherent states. The proof is based on two well-established notions: the Choi-Jamiolkowski state-channel duality  (see, e.g., \cite{Holevo10}) and the partial transpose \cite{Peres}.
 The state-channel duality is a standard tool to study the property of quantum channels whereas the partial transpose plays a central role in the theory of entanglement. Thanks to these reliable basics we can directly observe that the problem of the quantum benchmark is a type of separability problems on the quantum channel.  We also apply the present method to give a quantum-domain criterion associated with a set of experimentally measured fidelities.

We use a standard notation to denote the coherent state with the complex amplitude $\alpha$ by $\ket{\alpha}$ and the number state with the photon number $n$ by $\ket{n}$. The coherent state is expanded in the number basis as $\ket{ \alpha}= e^{-|\alpha |^2 /2} \sum_{n=0}^\infty \alpha ^n \ket{ n} /\sqrt{n!}$. When we work on the state with two modes, we call the first system $A$ and the second system $B$.

Let us define the average fidelity of a physical process $\mathcal E$ for the transformation task on the coherent states $\{|\sqrt N \alpha\rangle \} \to \{\ket{\sqrt \eta \alpha }\} $ with $N, \eta > 0$ by 
\begin{eqnarray}
F_{N, \eta, \lambda} (\mathcal E) &:=&  \int p_\lambda( \alpha )\bra{\sqrt \eta \alpha }\mathcal E \Big(|\sqrt N \alpha \rangle \langle \sqrt N \alpha| \Big) \ket{\sqrt\eta \alpha} d^2 \alpha  \nonumber \\ \label{eq1} 
\end{eqnarray}
where  the prior distribution of a symmetric Gaussian function with an inverse width of $\lambda >0 $ is given by \begin{eqnarray}
p_\lambda( \alpha ) :=  \frac{\lambda }{\pi} \exp (- \lambda |\alpha |^2 ).\label{eq2}\end{eqnarray} It reproduces the uniform distribution in the limit $\lambda \to 0$. In the first proof \cite{Ham05}, the unit-gain transformation $\{\ket{\alpha }\} \to \{\ket{\alpha }\}$ was considered so as to establish a benchmark for the channel that is expected to retrieve input states without disturbance, such as the action of ideal quantum teleportation and quantum memory.  
The factor $N$ was introduced to consider a type of state estimation from $N$-copies of the coherent states $\ket{\alpha } ^{\otimes N}$ in Ref. \cite{Nis07} while the factor $\eta$ was introduced to consider the effect of loss and amplification in Ref. \cite{namiki07}.

  \textit{Quantum benchmark fidelity.---}
The quantum benchmark fidelity for above transformation task is defined by the maximum of the fidelity in Eq. (\ref{eq1}) with respect to the optimization of the quantum channel $\mathcal E$ over EB channels and shown to be \cite{Ham05,namiki07,Nis07}
\begin{eqnarray}
\sup_{\mathcal E \in EB} F_{N,\eta, \lambda} (\mathcal  E) = {\frac{ N + \lambda }{ N+ \lambda + \eta  } }= : F_C (N,\eta, \lambda )  
    \label{main} 
\end{eqnarray} where $EB$ stands for the set of EB channels. 
  Since  we can verify the relation 
  $F_{N, \eta, \lambda}  = F_{\frac{N}{\eta },1, \frac{\lambda}{\eta }}, = F_{1, \frac{\eta }{N}, \frac{\lambda}{N} }$  from Eqs. (\ref{eq1}) and (\ref{eq2}), 
 it is sufficient to show the relation of Eq. (\ref{main}) either case of $\eta = 1$ \cite{Nis07} or case of $N= 1$ \cite{namiki07}. In the following we prove Eq. (\ref{main}) with $\eta = 1 $. The central idea for the present proof is to make a connection between the fidelity and a two-mode squeezed state via a sort of the state-channel duality. Then, the problem turns out to be a problem to find the maximum expectation value of an observable without entanglement, which can be solved by using the notion of the partial transpose. 
 
 \textit{Proof.--- }
Let us consider the following integration with the parameters $s, \kappa  \ge 0 $, and $0\le  \xi < 1 $, 
\begin{eqnarray}
 J_{ \mathcal E}(s, \kappa ,\xi ) 
&:=& \int  {d}^2 \alpha  p_s(\alpha ) \bra{  \alpha}_A \bra{\kappa \alpha^*}_B \mathcal\nonumber\\ & &  \mathcal E_A \otimes I_B \left( \ket{\psi_\xi}\bra{\psi_\xi} \right ) \ket{\kappa \alpha^*}_B
 \ket{ \alpha}_A    \label{start}
\end{eqnarray}
where  $\ket{\psi_\xi}= \sqrt{1-\xi ^2} \sum_{n=0}^\infty \xi^n\ket{n}\ket{n}$ is the two-mode squeezed state and  $I$ represents the identity process.  Using the relation $\bra{\alpha }\ket{\psi_\xi} = \sqrt{1-\xi ^2} e^{-(1-\xi ^2 )|\alpha | ^2 /2} \ket{\xi \alpha ^* }$ we can verify the following identity: \begin{eqnarray}
 J_{ \mathcal E}(s,\kappa ,\xi )   &=& \frac{s(1-\xi^2)}{\lambda}
 F_{N,1, \lambda} (\mathcal E ) \label{jj}
\end{eqnarray}
where the parameters are connected as 
\begin{eqnarray}
\lambda &=& s+ (1-\xi^2)\kappa ^2,   \label{lam}\\
\sqrt N &=& \kappa \xi. \label{n}
\end{eqnarray}
In order to find an upper bound of the fidelity we consider an upper bound of $J_{\mathcal E}$. 
If $\mathcal E$ is a MP scheme, $\rho_{\mathcal E} := \mathcal E \otimes I \left( \ket{\psi_\xi}\bra{\psi_\xi} \right )$ is a separable state \cite{16}. Then, there exists a separable state, say $\rho_{\mathcal E}=  \rho_{\mathcal E}^\star$, corresponding to the optimal MP scheme that 
 maximizes $ J_{\mathcal E}$, i.e., $J_{\mathcal E}( \rho_{\mathcal E}^\star )= \sup_{\mathcal E \in EB}  J_{ \mathcal E}$. This implies that $J_{\mathcal E}( \rho_{\mathcal E}^\star )= \sup_{\mathcal E \in EB}  J_{ \mathcal E}$  is bounded above by the maximum of $  J_{\mathcal E}( \rho_{\mathcal E} )$ when $ \rho_{\mathcal E}$ is optimized over the set of separable states,   namely,  the following inequality holds,  
\begin{eqnarray}
 \sup_{\mathcal E \in EB}  J_{ \mathcal E}(s,\kappa ,\xi)  &\le& \max_{\rho \in Sep.} \tr \left[ \rho  M  
\right] , \nonumber 
\end{eqnarray}
where $Sep.$ represents the set of separable states and 
\begin{eqnarray}
M:= \int p_s(\alpha) \ket{ \alpha  }\bra{ \alpha  } \otimes \ket{\kappa  \alpha ^*  }\bra{\kappa \alpha ^* }  d^2\alpha .  \nonumber
\end{eqnarray}
 Note that the maximum over separable states can be achieved by a product state and that the optimization over product states is equivalent to the optimization over their partial transpose. Hence, for any $\rho \in Sep.$, we can verify $ \tr [ \rho  M  ] \le \max_{\phi,\varphi }\tr M \ketbra{\phi}{\phi} \otimes \ketbra{\varphi}{\varphi} = \max_{\phi,\varphi }\tr M \Gamma [ \ketbra{\phi}{\phi} \otimes \ketbra{\varphi}{\varphi}] = \max_{\psi, \varphi} \tr \Gamma [M]  \ketbra{\phi}{\phi} \otimes \ketbra{\varphi}{\varphi} $ where $\Gamma [ \cdot]$ denotes the partial transposition map. This implies 
\begin{eqnarray} \sup_{\mathcal E \in EB}  J_{ \mathcal E}(s,\kappa ,\xi ) &\le& \max_{\rho \in Sep.}  \tr   \rho \Gamma [M ] 
 \le   \| \Gamma [ M ] \|  \label{ff} \end{eqnarray}
where the last inequality comes from the fact that the maximum over separable states is no larger than the maximum over all physical states  and $\| \cdot \|:= \max_{\braket{u}{u}=1} \bra{ u} \cdot \ket{u} $ denotes the maximum eigenvalue.
 Since the transpose of the coherent state with respect to the number basis acts as a phase conjugation, 
  by taking the replacement $\ket{\kappa  \alpha ^*  }\bra{\kappa  \alpha ^*  } \to \ket{\kappa   \alpha  }\bra{ \kappa   \alpha}  $ on $M$ we have
 $\Gamma [M ] = \int p_s(\alpha) \ket{ \alpha  }\bra{ \alpha  } \otimes \ket{\kappa  \alpha  }\bra{\kappa \alpha }   d^2\alpha$. 
 By using the beam-splitter transformation $\hat V | \sqrt{1 + \kappa ^2 } \alpha  \rangle  \ket{0} = \ket{ \alpha}\ket{  { \kappa   } \alpha} $ we can write  
\begin{eqnarray}  \|  \Gamma [M]  \| &= & \| \hat V^\dagger \Gamma [M] \hat V \|\nonumber \\   & =& \left\|  \int p_s( \alpha )  |\sqrt{1+\kappa  ^2 }\alpha \rangle\langle  \sqrt{1+\kappa   ^2 } \alpha  |  \otimes |0\rangle \langle0 | d^2 \alpha \right\|\nonumber \\ &=& \left\| T \left( \frac{1+ \kappa^2}{s}\right ) \otimes \ketbra{0}{0} \right\| =  \frac{s }{s+1 +\kappa    ^2} \label{ongm}
  \end{eqnarray} 
  where  
\begin{eqnarray}
 T( \bar n )  &:= & {\frac{1 }{ 1+ \bar n   }} \sum_{n =0}^{\infty} \left( \frac{\bar n }{ 1+ \bar n  } \right)^{n }    |n  \rangle \langle n  |  \nonumber 
\end{eqnarray}
  is the   thermal state with the mean photon number $\bar n $. 
Equations (\ref{ff}) and (\ref{ongm}) lead to 
\begin{eqnarray}
 \sup_{\mathcal E \in EB}  J_{ \mathcal E}(s,\kappa ,\xi ) 
&\le & \frac{s }{s+1+ \kappa ^2 }. \nonumber 
\end{eqnarray}
Using this relation and Eqs.  (\ref{jj}), (\ref{lam}) and (\ref{n}),  we have 
\begin{eqnarray}
 \sup_{\mathcal E \in EB}  F_{N,1, \lambda } (\mathcal E ) 
&\le& \frac{\lambda}{ (1-\xi^2)} \frac{1}{N+ \lambda +1 }  \label{sonamae} . 
\end{eqnarray}
From the condition $s\ge 0 $ with Eqs. (\ref{lam}) and (\ref{n}),  we have \begin{eqnarray}
\frac{\lambda}{1-\xi ^2} \le N+ \lambda . \label{scondition}
\end{eqnarray}
From  
 Eqs. (\ref{sonamae}) and (\ref{scondition}), we  obtain the upper bound   
\begin{eqnarray}
 \sup_{\mathcal E \in EB}  F_{N,1, \lambda } (\mathcal E ) &\le&   \frac{N+\lambda }{N+ \lambda + 1  } =  F_C(N, 1, \lambda). 
\end{eqnarray} 
This bound can be achieved by the EB channel 
$\mathcal E_{EB} (\rho ):= \frac{1}{\pi} \int \bra{\alpha }\rho \ket{ \alpha} \ket{\frac{\sqrt{ N} \alpha }{N +\lambda} }\bra{\frac{\sqrt{  N}\alpha }{N +\lambda} } d^2 \alpha$. We thus have  $\sup_{\mathcal E \in EB}  F_{N,1, \lambda } (\mathcal E ) \ge F_{N,1,\lambda}(\mathcal E _{EB})=  F_C(N, 1, \lambda)$. 
This concludes Eq. (\ref{main}) with $\eta =1$.\hfill$\blacksquare$

It is well-known that the inseparability of two-mode Gaussian states can be characterized by using the standard form of the covariant matrices of the Gaussian states under the local Gaussian unitary operators \cite{Duan00}. Similarly, one-mode Gaussian channels can be described by a pair of $2$-by-$2$ matrices that determines the transformation of the covariant matrices and are classified into a few standard forms under the suitable unitary operations before-and-after the channel \cite{Hol08}. Two of the standard forms are relevant to quantum domain channels. In both forms one can find a proper set of the parameters $(N, \eta , \lambda)$ so that the classical limit fidelity is surpassed if the given channel is in quantum domain \cite{namiki07}. In this sense, the output-target fidelity with the Gaussian distributed set of coherent states is capable of detecting any one-mode Gaussian channels in quantum domain. 

In experiments we usually obtain a finite set of measured fidelities. The set of data is not enough to directly calculate the integration in Eq. (\ref{eq1}), and $F (\mathcal E )$ is estimated by  using additional assumptions.
It is better if one can check a quantum domain criterion directly associated with the set of measured fidelities without  additional assumptions. 
 In the following we present a general theorem to produce a quantum domain criterion associated with a given set of measured fidelities. The proof of this theorem is essentially the same as above proof. It is remarkable that the criterion can be generated by a simple calculation of a maximum eigenvalue. %

\textit{In-situ generation of a quantum-domain condition.---} 
 Let us write a set of input states $\{\ket{\psi_i} \}$, a set of target states $\{\ket{\psi_i '} \} $, and a prior probability distribution $\{p_i\}$ with $\sum_i p_i=1 $.   We can show that the following theorem holds: A process $\mathcal E$  is in quantum domain if 
\begin{eqnarray}
\bar F [\mathcal E ;  p_i; \psi_i  \to \psi_i '  ] > d \left\| \sum_i p_i  \ketbra{\psi_i '}{\psi_i '}
 \otimes  \ketbra{\psi_i }{\psi_i }  \right\| ,  \label{QDC}
\end{eqnarray}
where the average fidelity is given by \begin{eqnarray}
 \bar F [\mathcal E ;  p_i; \psi_i  \to \psi_i '  ] := \sum_i p_i \bra{\psi_i '} \mathcal E (\ketbra{\psi_i }{\psi_i }) \ket{\psi_i '} ,   \nonumber 
 \end{eqnarray} %
and $d$ is the dimension of the Hilbert space spanned by the set of input states $\{\ket{\psi_i} \}$. Note that the experiment determines the  set of the fidelities $ \{  \bra{\psi_i '} \mathcal E (\ketbra{\psi_i }{\psi_i }) \ket{\psi_i '} \} $ whereas the  choice of the  probability distribution  $\{p_i\}$ is arbitrary.

 \textit{Proof.--- }
 Let 
  $\{\ket{u_k}\}_{k=0,1,2, \cdots , d-1 }$ be an orthonormal basis of the $d$-dimensional Hilbert space. We define the maximally entangled state of a two-$d$ level system by $ \ket{\Phi_d }:= \sum_{k=0}^{d-1} \ket{u_k}\ket{u_k} /\sqrt{d} $. We also define the complex conjugation of the $d$-dimensional state by $\ket{\psi^*  }:= \sum_k \braket{\psi}{u_k} \ket{{u_k}} = \sqrt d \bra{\psi} \ket{\Phi_d } $. Then, we can write 
\begin{eqnarray}
&& \bar F [\mathcal E ;  p_i; \psi_i  \to \psi_i '  ] \nonumber \\
 &=& d \sum_i p_i \bra{\psi_i '}_A \bra{{\psi_i ^*}}_B \mathcal E_A \otimes I_B (\ketbra{\Phi_d }{\Phi_d }  )    \ket{\psi_i^*  }_B   \ket{\psi_i '}_A \nonumber \\
  &=& d \tr [  M \rho_{\mathcal E} ]
   \label{lllf} 
\end{eqnarray}
where we write 
$M = \sum_i p_i (\ketbra{\psi_i '}{\psi_i '})_A \otimes ( \ketbra{{\psi_i^* }}{\psi_i^* })_B $ and $\rho_{\mathcal E} = \mathcal E_A \otimes I_B (\ketbra{\Phi_d }{\Phi_d }  )$. The state $\rho_{\mathcal E}$ is the standard Choi-Jamiolkowski isomorphism. In the continuous-variable case, we have used a two-mode squeezed state instead of an unnormalizable maximally entangled state \cite{Holevo10}.


If the process $ \mathcal E$ is a MP scheme, $\rho_{\mathcal E} = \mathcal E_A \otimes I_B (\ketbra{\Phi_d }{\Phi_d }  )$ belongs to the set of separable states \cite{16}. Hence, the maximum of the average fidelity over all MP schemes is bounded above by the maximum of the final expression of Eq. (\ref{lllf}) achieved by the optimization of the state $\rho_{\mathcal E}$ over separable states. This implies 
\begin{eqnarray}
\max_{\mathcal E \in EB} \bar F [\mathcal E ;  p_i; \psi_i  \to \psi_i '  ] \le d \max_{\rho  \in Sep.} \tr [ M \rho ] .   \label{17}
\end{eqnarray}
  Since the optimization over separable states can be converted into the optimization over their partial transpose, we have  
\begin{eqnarray}
 \max_{\rho  \in Sep.} \tr [ M  \rho ] &=&  \max_{\rho  \in Sep.} \tr [ \Gamma [M ] \rho ]  \label{18} 
\end{eqnarray} where $\Gamma$ stands for the partial transposition map again.
Since the maximum over separable states is bounded above by the maximum over all physical states, we have
\begin{eqnarray}
 \max_{\rho  \in Sep.} \tr [ \Gamma[  M ] \rho ] 
  &\le&  \max_{\rho } \tr [ \Gamma [M ] \rho ] 
   =  \| \Gamma [M ]  \|.   \label{19} 
\end{eqnarray}
When we choose the partial transposition of the second system with respect to   
the basis $\{\ket{u_k}\} $, we have 
\begin{eqnarray}
\Gamma [M]=   \sum_i p_i  \ketbra{\psi_i '}{\psi_i '}
 \otimes  \ketbra{\psi_i }{\psi_i }. \label{GammaM}
\end{eqnarray}
Concatenating Eqs. (\ref{17})-(\ref{19}) and (\ref{GammaM}) we can see that the maximum fidelity  over all MP schemes is bounded above by right hand side of Eq. (\ref{QDC}). Hence, if a quantum channel provides the fidelity higher than this limit, it is incompatible with any classical MP scheme. \hfill$\blacksquare$

Consequently, if Ineqs. (\ref{17}) and (\ref{19}) are tight we can immediately obtain the classical limit just by  the calculation of the maximal eigenvalue of the operator in right hand side of Eq. (\ref{QDC}). This is the case for the following example.

\textit{Example.--- }
Let us consider the uniform set of input states over the $d$-dimensional Hilbert space and transformation task of a unitary map by setting the target state $ \ket{\psi'} = U \ket{\psi }$ for any input $\ket{\psi}$. 
 In this case it is well-known  that  the classical limit fidelity is  given by  \cite{Pop94,Mass95,Horo99,Brass99,Fuc03} 
\begin{eqnarray}
\bar F_c^{(d)}&:=& \max_{\mathcal E \in EB }  \int d\psi \bra{\psi} U^\dagger \mathcal E ( \ketbra{\psi}{\psi} ) U \ket{\psi} \nonumber \\
& =& \max_{\mathcal E \in EB }  \int d\psi \bra{\psi} \mathcal E ( \ketbra{\psi}{\psi} ) \ket{\psi} = \frac{2}{d+1}. \nonumber  
\end{eqnarray}
 where $\int d\psi$ denotes the Haar measure and the second equation comes from the fact that the total action of an EB channel followed by a unitary map can be described by a single EB channel. Hence, it is sufficient to consider the case that the task is the identity transformation, i.e., $\ket{\psi  '} =\ket{\psi}$. 
 For the uniform ensemble of input states, the state of Eq. (\ref{GammaM}) becomes the so-called Werner state \cite{wer89, Voll01}, and is decomposed into  
$\Gamma [M]=   \int d\psi \ketbra{\psi}{\psi}  \otimes  \ketbra{\psi}{\psi} =( \openone   + f )/[d(d+1)]  $, 
where $f:= \sum_{i,j} \ketbra{u_i}{u_j } \otimes \ketbra{ u_j}{ u_i}$ is the   flip operator. Hence, we have $\| \Gamma [M] \| =\frac{2}{d(d+1)} $, 
and obtain the inequality  $\max_{\mathcal E  \in EB}\bar F \le  d \| \Gamma [M] \|  = \frac{2}{d+1} =\bar F_c^{(d)}$  through Eqs. (\ref{17}), (\ref{18}), and (\ref{19}). The inequality is saturated by the EB channel $\mathcal E _{EB}( \rho )= \sum_j U \ketbra{u_j}{u_j} \rho \ketbra{u_j}{u_j} U^\dagger $. This can be confirmed by the following equations: 
$ \bar F = \int d\psi \bra{\psi} U^\dagger \mathcal E_{EB} ( \ketbra{\psi}{\psi} ) U \ket{\psi}=   \tr  [\sum_{j} \ketbra{u_j}{u_j } \otimes  \ketbra{u_j}{u_j }    (  \openone + d \ketbra{\Phi_d}{ \Phi_d } ) ]/[d(d+1)] = \frac{2}{d+1}$ 
 where
we used the relation $ \int d\psi \ketbra{\psi}{\psi} \otimes \ketbra{\psi^* }{\psi ^* }  = (\openone  + d \ketbra{\Phi_d}{ \Phi_d })/[d(d+1)] $ in the second line (see, e.g., \cite{Voll01}). Hence we obtain the tight classical limit. In the previous approaches \cite{Pop94,Mass95,Horo99,Brass99,Fuc03}, the problem is treated as a type of state estimation in Refs. \cite{Pop94,Mass95} and is also connected to a limit of optimal cloning in \cite{Brass99} whereas it is addressed as separability problems in Refs. \cite{Horo99,Fuc03}. Our approach is somehow close to the approach of Ref. \cite{Horo99} in the sense that the maximally entangled state plays a central role.

In conclusion, we have presented an alternative proof of the quantum benchmark fidelity with respect to a Gaussian distributed set of coherent states. The main idea of proof is to use a sort of the state-channel duality to associate the average fidelity to the two-mode squeezed state. Then, the partial transpose is utilized to make the bound on the fidelity as a separability problem. Based on this method we have also presented a general theorem to produce a quantum-domain criterion associated with a set of measured fidelities.  The theorem can be utilized in a wide class of experiments. The present method  would be useful to further comprehend the property of quantum channels.
   
R.N. acknowledges support from JSPS.

\end{document}